\documentclass[aps,prb,twocolumn,showpacs,preprintnumbers,amsmath,amssymb]{revtex4}
\usepackage[dvips]{graphics}

\usepackage{graphicx}
\usepackage{dcolumn}
\usepackage{bm}

\begin{document}
\preprint{APS/123-QED}

\title{Magnetic structure of CeRhIn$_{5}$ as a function of pressure and temperature}
\author{A. Llobet$^{1,*}$, J. S. Gardner$^{2,\dag}$, E. G. Moshopoulou$^{3}$, J.-M. Mignot$^{4}$, M. Nicklas$^{1,\ddag}$, W. Bao$^{1}$, N.O. Moreno$^{1}$, P.G. Pagliuso$^{1,\S}$, I. N. Goncharenko$^{4}$, J. L. Sarrao$^{1}$,
 and J. D. Thompson$^{1}$}
\affiliation{$^{1}$Los Alamos National
Laboratory, Los Alamos, NM 87545\\$^{2}$NRC Canada, NPMR, Chalk
River Laboratories, Chalk River, Ontario,Canada K0J
1J0\\$^{3}$National Center for Scientific Research ``Demokritos",
Institute of Materials Sciences, 15310 Agia Paraskevi,
Greece\\$^{4}$Laboratoire L\'{e}on Brillouin (CEA-CNRS),
CEA/Saclay, 91191 Gif-sur-Yvette Cedex, France}

\date{\today}

\begin{abstract}
We report magnetic neutron-diffraction and electrical resistivity
studies on single crystals of the heavy-fermion antiferromagnet
CeRhIn$_{5}$ at pressures up to 2.3~GPa. These experiments show
that the staggered moment of Ce and the incommensurate magnetic
structure change weakly with applied pressure up to 1.63~GPa,
where resistivity, specific heat and NQR measurements confirm the
presence of bulk superconductivity. This work places new
constraints on an interpretation of the relationship between
antiferromagnetism and unconventional superconductivity in
CeRhIn$_{5}$.

\end{abstract}

 \maketitle
\section{Introduction}
Heavy fermion (HF) materials provide an excellent opportunity to
investigate the interaction between magnetism and unconventional
superconductivity (SC). In most HF compounds the magnetic
interactions are governed by the hybridization of the \textit{f}
electrons and the conduction electrons. This leads to competition
between the tendency to order magnetically, favored by the RKKY
(Ruderman-Kittel-Kasuya-Yosida) indirect exchange interaction, and
the tendency to have a spin-singlet ground state mediated by the
Kondo interaction. In these systems, there is increasing
experimental and theoretical evidence that antiferromagnetic (AFM)
spin fluctuations mediate Cooper pairing and that anisotropic SC
appears in the vicinity of a quantum-critical
point~\cite{MathurND:Magmsh, SatoNK:Strcbl, FiskZ:Hea-el,
MiyakeK:Spi-fl, monthB,ColemanP:Whaift}. Several families of HF
compounds are known where SC does coexist with weak magnetic order
(e.g.,UPt$_{3}$, URu$_{2}$Si$_{2}$,
UNi$_{2}$Al$_{3}$\cite{HeffnerRH:Heafs.}). However most of
Ce-based heavy fermion superconductors (HFS)~(CeIn$_{3}$\cite{
WalkerIR:norssC,MorinP:Magsup},
CeCu$_{2}$Ge$_{2}$~\cite{JaccardD:Magshf,JaccardD:Preihf},
CePd$_{2}$Si$_{2}$~\cite{ThompsonJD:EffptN,
GroscheFM:MagspC},CeRh$_{2}$Si$_{2}$~\cite{MovshovichR:Suph-f})
display an AFM ground state at ambient pressure and superconduct
when external pressure is applied and \textit{T}$_{\textrm{N}}$ is
driven to 0~K.
\par A new family of Ce-based compounds: Ce\textit{M}In$_{5}$ (\textit{M}=Co,Ir,Rh) with
Sommerfeld coefficients ($\gamma$) of 1000, 750 and 380~mJ
mol$^{-1}$K$^{-2 }$ respectively, has recently been added to the
list of
HFS~\cite{HeggerH:Pre-in,PetrovicC:newh-f,PetrovicC:Hea-fe,ZhengG-q:Unisdu,MovshovichR:UncsC5,FisherRA:SpehC5}.
The most notable properties in this series include:
ambient-pressure magnetic order (\textit{T}$_{\textrm{N}}$ =
3.8~K) and pressure-induced SC (\textit{T}$_{\textrm{C}}$ = 2.1~K
at P=1.63~GPa) in CeRhIn$_{5}$~\cite{HeggerH:Pre-in},
unconventional ambient-pressure SC in both CeIrIn$_{5}$
(\textit{T}$_{\textrm{C}}$ = 0.4~K)~\cite{PetrovicC:Hea-fe} and
CeCoIn$_{5}$ (\textit{T}$_{\textrm{C}}$ =
2.3~K)~\cite{PetrovicC:Hea-fe} and coexistence of SC and AFM in
solid solutions of CeRh$_{1-x}$\textit{A}$_{x}$In$_{5}$
(\textit{A}=Ir,Co)~\cite{PagliusoPG:CoemsC,ZapfVS:CoesaC}. The
HoCoGa$_{5}$-type crystal structure is common to these compounds
(a=b$\simeq~$4.66~\AA~and
c$\simeq$~7.51~\AA)\cite{MoshopoulouEG:Crygis} and consists of
alternating layers of CeIn$_{3}$ and \textit{M}In$_{2}$ stacked
sequentially along the [001] direction.\par Here we focus on
CeRhIn$_5$ which orders at ambient pressure in an incommensurate
AFM helical structure~\cite{CurroNJ:Evismo} with a wave vector
\textit{q}$_{m}$=(0.5,0.5,0.297)\cite{bao}. For P$\geq$1.5~GPa,
the AFM ground state is
 suppressed and bulk SC is observed below
\textit{T}$_{\textrm{C}}$=~2.2~K~\cite{HeggerH:Pre-in,
FisherRA:SpehC5}. $^{115}$In-NQR studies on CeRhIn$_{5}$ initially
reported a decrease with applied pressure of the internal magnetic
field (\textit{H$_{int}$}) at the In(1)site (located in the
CeIn$_{3}$ layer) and the authors attributed this decrease either
to a reduction of the ordered moment of Ce or its rotation from
the \textit{a-b} plane with pressure\cite{MitoT:SysNst,
MitoT:Pre-in}. Recent NQR studies found that AFM coexists
homogeneously with SC at a microscopic level in CeRhIn$_{5}$ and
estimate that if the ordered moment is reduced with pressure, its
value at 1.75~GPa, where AFM coexists with bulk SC, is at most
5~\%\ of its ambient pressure value\cite{MitoT:cond}. On the other
hand, specific heat measurements on CeRhIn$_{5}$ show that the
entropy below \textit{T}$_{\textrm{N}}$ decreases by about 20~\%\
as pressure is raised from 0.3 to 1.32~GPa \cite{FisherRA:SpehC5}
which appears to be inconsistent with the reduction of the ordered
moment suggested by NQR results. Neutron diffraction studies are
required for an unambiguous determination of the pressure
evolution of the ordered moment and magnetic structure.\par An
initial neutron diffraction study under hydrostatic pressure
(P$<$0.4~GPa) on CeRhIn$_{5}$~\cite{WeiBao:pressure} revealed no
changes in the staggered moment nor \textit{T}$_{\textrm{N}}$
within the error bars of the experiment and reported a slight
reduction in the incommensurate wave number. Consistent with this
first study, more recent neutron diffraction measurements find
essentially no change in the magnetic structure at low
pressures~\cite{MajumdarS:pre}. However, around 1~GPa, Majumdar
\textit{et al.} reported a marked change in the wave vector from
(0.5,0.5,0.298) to (0.5,0.5,0.396) and a 20~\%\ reduction of the
magnetic moment. They also reported no evidence for AFM order
above 1.3~GPa which contrasts with resistivity, specific heat and
NQR observations \cite{HeggerH:Pre-in, FisherRA:SpehC5,
KawasakiS:Pre-te, MitoT:cond, MitoT:SysNst, MitoT:Pre-in} and
tentatively ascribe the change in the magnetic structure to a
change in the electronic structure under pressure. However, de
Haas-van Alphen results show that the topology of the Fermi
surface does not change at pressures below
2.0~GPa~\cite{ShishidoH:Evolpre} and that there is a steep
increase in the cyclotron mass only above 1.6~GPa when SC
develops. Additional X-ray diffraction studies confirm that the
CeRhIn$_{5}$ crystal structure, except for a small decrease in the
cell volume, remains unchanged for pressures up to
2.0~GPa~\cite{kumar}. In this work, we have extended the pressure
range over which electrical resistivity and neutron diffraction
measurements have been performed on CeRhIn$_{5}$ in order to
investigate the effect of pressure on the superconducting and
magnetic transition temperatures and the evolution of the magnetic
structure as it approaches and exceeds the critical pressure where
the two phases meet.
\section{Experimental details}
Single crystals of CeRhIn$_{5}$ were grown using the In flux
technique~\cite{mosh, pagliuso_co}. Four-probe AC resistivity
measurements, with the current flowing in the tetragonal basal
plane, were made on bar-shaped single crystals. A clamp-type cell
generated hydrostatic pressures to 2.3 GPa for resistivity
measurements using silicon oil as the pressure medium.\par Neutron
diffraction experiments were carried out at the C5 and N5
spectrometers at the NRU reactor, Chalk River Laboratories (CRL)
as well as at the 6T2 lifting detector diffractometer at
Laboratoire L\'{e}on Brillouin,Saclay (LLB). A clamp-type Cu-Be
cell~\cite{ThompsonJD:Low-te} was used in experiments performed at
CRL with Fluorinert-75 as the pressure medium to generate up to
1.8~GPa. Bar-shaped single crystals
(1.3~$\times$~1.3~$\times$~10~mm) were used to reduce neutron
absorption by In and Rh nuclei at CRL. The longest dimension of
the crystals was along the (1$\overline{1}$0) crystallographic
axis. The scattering plane was defined to be the (hhl). In this
set-up, the applied pressure was determined, within $\pm$ 0.1~GPa,
by measuring the lattice parameters of a graphite crystal placed
behind the sample inside the cell at low temperature. Neutron
beams with incident energy of E$_{i}=$~35 meV were produced from a
Ge(113) or Be(002) monochromator. Pyrolytic graphite (PG) filters
with approximate thickness of 10~cm were placed in the scattered
beam to reduce higher order reflections and occasionally a
pyrolytic graphite analyzer was used to improve the signal to
noise ratio.\par
\begin{figure}[tb]
 \vspace{0.0cm}
\begin{center}
\includegraphics[width=0.5\textwidth]{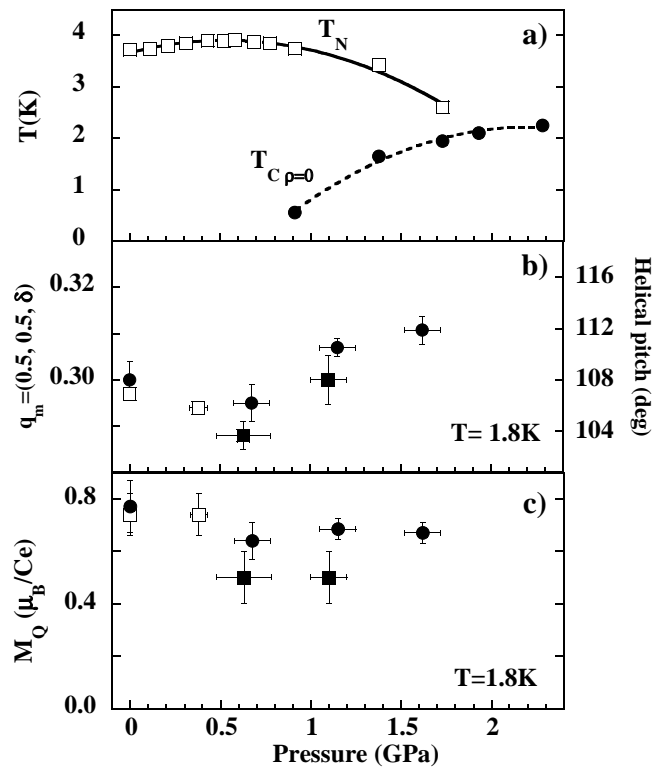}
\caption{a) Temperature-pressure
  phase diagram for CeRhIn$_{5}$ determined by $\rho$(T) measurements. Open squares correspond
  to the N$\acute{e}$el temperature and solid circles to the temperature at which the resistivity drops to zero. The lines are guides to the eye.
  b) Pressure evolution of the antiferromagnetic helical
structure characterized by the propagation vector
\textit{q}$_{m}$=(0.5,0.5,$\delta$). c) Pressure evolution of the
estimated Ce staggered moment at T=1.85~K of CeRhIn$_{5}$. Filled
circles correspond to measurements performed at CRL., filled
squares correspond to measurements performed at LLB Saclay and
empty squares correspond to data reported by Bao \textit{et
al}.~\cite{bao, WeiBao:pressure} }\label{fig1a}
\end{center}
\vspace{0.0cm}
\end{figure}
At LLB, a gasketed sapphire anvil cell was used with a mixture of
methanol and ethanol as the pressure transmitting medium for
experiments at 0.63 and 1.1~GPa. Samples, with dimensions
1.3~$\times$~1.3~$\times$~0.2~mm, were aligned with the [001]
crystallographic direction (shorted dimension) vertical . A thin
layer of ruby powder was placed on the inner surface of the anvil
in order to measure the applied pressure at room temperature using
the standard ruby fluorescence technique. This was performed
before and after thermal cycling to ensure that pressure was
constant throughout the experiment. This technique allows
determining the pressure at low temperatures within $\pm$
0.15~GPa. Neutron beams of E$_{i}=$ 14.81~meV, were produced using
a PG(002) monochromator. In both laboratories a top loading
He-flow cryostat was used to cool down the pressure cell and
sample. Results reported below were obtained on several different
single crystals, in different pressure environments and at two
neutron sources. The consistency of these results substantiate
conclusions drawn from them.

\section{Results and discussion}

We measured electrical resistivity ($\rho$) on CeRhIn$_{5}$ single
crystal at different applied pressures and temperatures from
300~mK to room temperature. This crystal has a resistivity ratio
$\rho$(295~K)/$\rho$(1.5~K)=~280 that is about two times higher
than that in a crystal studied by Hegger \textit{et
al.}\cite{HeggerH:Pre-in}. The pressure-temperature (P-T) phase
diagram constructed from these $\rho$(T) measurements is shown in
Fig.\ref{fig1a}a. Our new results show that the magnetic to
non-magnetic transition is smooth and reveals the existence of a
large pressure region of coexisting long-range magnetic order and
SC (0.9~GPa$\leq$P$\leq$1.75~GPa). There is a slight increase of
\textit{T}$_{\textrm{N}}$ with pressure up to about 0.8~GPa and
for pressures above this value \textit{T}$_{\textrm{N}}$ decreases
and a SC ground state develops. This phase diagram is fully
consistent with that determined by specific heat
\cite{FisherRA:SpehC5}~and NQR \cite{
MitoT:SysNst,MitoT:Pre-in,MitoT:cond} and departs from initially
reported results \cite{HeggerH:Pre-in}.\par
\begin{figure}[tb]
 \vspace{0.0cm}
\begin{center}
 \includegraphics[width=0.45\textwidth]{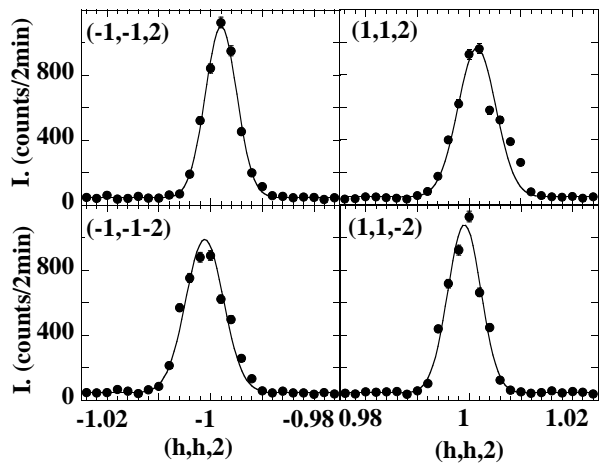} \caption{Elastic \textit{q} scans though
selected nuclear Bragg peaks at T=1.8~K and P=1.15~GPa.}
\label{fig1}
\end{center}
\vspace{0.0cm}
\end{figure}

\begin{figure}[tb]
 \vspace{0.0cm}
\begin{center}
 \includegraphics[width=0.45\textwidth]{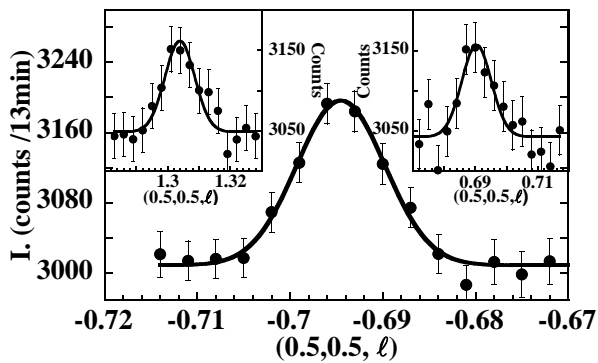} \caption{Elastic \textit{q} scans
around some of the magnetic peaks at T=1.8~K and P= 1.15~GPa.}
\label{fig2}
\end{center}
\vspace{0.0cm}
\end{figure}
\begin{figure}[tb]
 \vspace{0.0cm}
\begin{center}
 \includegraphics[width=0.45\textwidth]{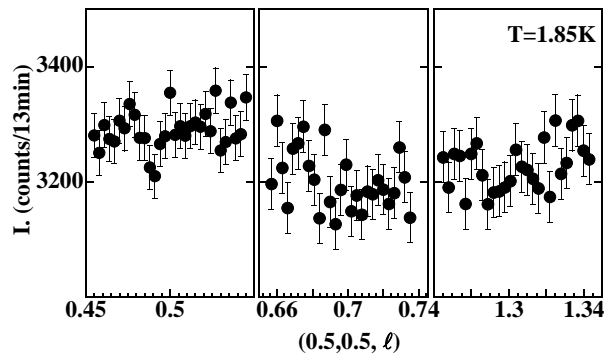}
 \caption{Selected elastic \textit{q} scans at T=1.8~K and
 P= 1.8~GPa not showing any evidence of magnetic scattering.}~\label{fig7}
\end{center}
\vspace{0.0cm}
\end{figure}

To determine the pressure evolution of the magnetic structure of
CeRhIn$_{5}$ and particularly the incommensurability parameter
($\delta$) of the magnetic structure (\textit{q}$_{m}$ =
(0.5,0.5,$\delta$)) special attention has been paid to the precise
alignment of the single crystal since $\delta$ depends critically
on it. For this reason, systematic checks have been performed
during the measurements using \{1,1,2\}, \{0,0,3\} and \{2,2,0\}
nuclear Bragg reflections. Fig.\ref{fig1} shows typical
\textit{q}-scans around a series of \{1,1,2\} reflections at
1.85~K and 1.15~GPa which attest to the quality of the crystal
alignment. When changing pressure, the cell and sample were warmed
to room temperature before the next pressure was applied. At each
pressure, \textit{q}-scans and rocking curves were measured at
magnetic and nuclear peaks. Several magnetic reflections,
including Friedel pairs, were measured to determine $\delta$ more
accurately. A set of representative magnetic Bragg peaks are shown
in Fig.\ref{fig2} for P=1.15~GPa and T=1.85~K. The absence of
other commensurate reflections, like (0.5,0.5,0.5), was also
systematically verified. From data such as shown in
Fig.\ref{fig2}, we obtain the pressure dependence of $\delta$
plotted in Fig.\ref{fig1a}b. Our results show that there is no
substantial change in the magnetic wave vector (0.5,0.5,$\delta$)
within the accuracy of these measurement up to pressures of
1.63~GPa. This is qualitatively different from the result reported
by Majumdar {\it et al.} \cite{MajumdarS:pre}. At 1.8~GPa, we do
not detect any evidence for magnetic scattering for temperatures
greater than 1.85~K as shown in Fig.\ref{fig7}. We speculate that
the lack of magnetic long range order at this pressure can be due
to the existence of a marginally higher pressure than 1.8~GPa
which would drive \textit{T}$_{\textrm{N}}$ close to our lowest
measuring temperature in which case the magnetic scattering would
be not observable above background scattering from the Be-Cu
pressure cell. The possibility that a dramatic change may occur in
the magnetic structure between 1.63~GPa and 1.8~GPa giving no
magnetic scattering along (0.5,0.5,$\ell$) for the $\ell$ interval
reported seems very unlikely but cannot be definitely ruled
out.\par
\begin{figure}[tb]
 \vspace{0.0cm}
\begin{center}
 \includegraphics[width=0.40\textwidth]{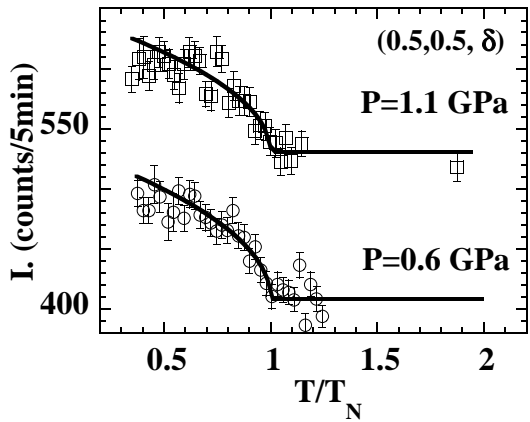} \caption{Temperature dependence of the
 $($0.5,0.5,$\delta)$
 Bragg peak intensity at P=0.6~GPa and 1.1~GPa (vertical offset added) and \textit{T}$_{\textrm{N}}$(0.6~GPa)~$\sim$~\textit{T}$_{\textrm{N}}$(1.1GPa). The solid lines
are fits to (1-T/\textit{T}$_{\textrm{N}})^{2\beta}$ with
$\beta$=~0.25.} \label{fig4}
\end{center}
\vspace{0.0cm}
\end{figure}

The temperature dependence of the $($0.5,0.5,$\delta)$ Bragg peak
intensity which corresponds to the magnetic order parameter
squared  is shown in Fig.\ref{fig4} for P=~0.6~GPa and 1.1~GPa. It
reveals that there is not a significant change in the development
of the magnetic order at pressures above and below the pressure
where SC starts developing. A tentative fit to
(1-T/T$_{N}$)$^{2\beta}$ showed better agreement when
$\beta$=~0.25 which is consistent with the results reported at
ambient pressure~\cite{WeiBao:Anit-d}.\par To determine the
magnetic moment at each pressure, magnetic Bragg peaks were
measured at 1.8~K with rocking scans at LLB-Saclay and with scans
such as those in Fig.\ref{fig2} at CRL. Magnetic cross-sections
are derived from integrated intensities with appropiate correction
for resolution \cite{CooperMJ:resfnd}. They are normalized to
nuclear Bragg peaks to yield values in absolute unit.
The theoretical cross-section for the AFM spiral model is
\cite{bao,jean:anna:artem}:
\begin{equation}\label{squires1}
    \sigma(q)=(\frac{\gamma r_{0}}{2})^{2}( M_{Q} )^{2}
    \frac{1}{4}|f(q)|^{2}(1+(\widehat{\mathbf{q}}\cdot
    \widehat{\mathbf{c}})^{2})
    \end{equation}
where $f(q)$ is the Ce$^{+3}$ magnetic form factor \cite{blume},
$(\frac{\gamma_{0}r_{0}}{2})^{2}$=~0.07265 barns/$\mu_{B}$ and
\textit{M$_{Q}$} is the staggered moment of Ce ion.
Fig.\ref{fig1a}c shows the staggered moment of Ce as a function of
applied pressure. The staggered magnetic moment of Ce at ambient
pressure, \textit{M$_{Q}$}=~(0.8~$\pm~$0.1)~$\mu_{B}$/Ce, which is
consistent with the previously reported value of
\textit{M$_{Q}$}=~(0.75~$\pm~$0.02)~$\mu_{B}$/Ce and is found to
be about 20\%\ smaller than the full moment obtained from crystal
field calculations, which estimate
\textit{M$_{Q}$}=0.92~$\mu_{B}$/Ce~\cite{cris}. We attribute the
smaller measured value of \textit{M$_{Q}$} to partial Kondo
compensation of the moment, an effect neglected in crystal field
calculations. Fig.\ref{fig1a}c also shows that there appears to be
a slight tendency for \textit{M$_{Q}$} to decrease with pressure
(less than 15$\%$ decrease at 1.63~GPa compared to ambient
pressure).
\par
\begin{figure}[tb]
\begin{center}
 \includegraphics[width=0.45\textwidth]{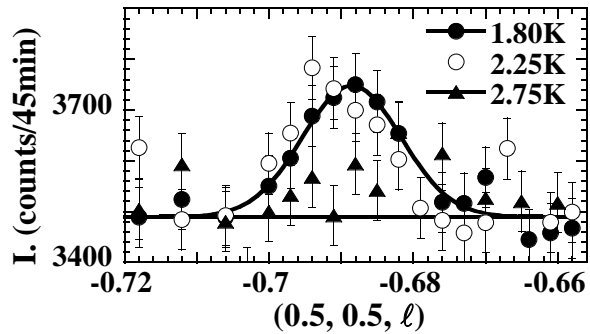}
 \caption{\textit{q}-scans around (0.5,0.5,1-$\delta$) magnetic reflection at 1.63~GPa
for different temperatures above and below
\textit{T}$_{\textrm{N}}\simeq$2.85~K using a pyrolytic graphite
analyzer to improve the signal to background ratio.} \label{fig6}
\end{center}
\end{figure}

An anomaly at T$_{?}$=2.8~K was found for 1.3$<$P$<$2.0~GPa in
earlier resistivity measurements on
CeRhIn$_{5}$\cite{HeggerH:Pre-in}. This resistivity anomaly is not
detected in the higher quality crystals used to construct the
phase diagram in Fig.\ref{fig1a}. We have measured
\textit{q}-scans around (0.5,0.5,$\delta$) at P=~1.63~GPa for
different temperatures (Fig.\ref{fig6}) and our results confirm
that magnetic long range AFM helical order disappears between 2.25
and 2.75~K which is very close to values of T$_{N}$ obtained from
resistivity measurements (T$_{N}$(1.6~GPa)=2.8~K).\par For the
incommensurate magnetic structure of CeRhIn$_{5}$, the internal
magnetic field sensed by $^{115}$In-NQR is given by
\textbf{H}$_{int}\propto$
\textit{A$_{ab}$}M$_{Q}$\{$\sin$(q$_{0}$\textit{z}),$\cos$(q$_{0}$\textit{z}),0\},
where $A_{ab}$ is the hyperfine coupling between the in-plane In
nucleus and each of its four Ce nearest neighbors,
\textit{M$_{Q}$} is the ordered moment and \textit{q}$_{0}$=
2$\pi\delta/$c \cite{CurroNJ:cond}. Our neutron diffraction
experiments show that $\delta$ and \textit{M$_{Q}$} change by at
most 10\%\ and 15\%\ respectively, as pressure is raised from
atmospheric to 1.63~GPa (Fig.\ref{fig1a}(b) and
Fig.\ref{fig1a}(c)). These relatively small changes in $\delta$
and \textit{M$_{Q}$} are unable by themselves to account for the
80\%\ reduction  of \textit{H$_{int}$} deduced by NQR
measurements. If the Ce moments acquire a component out of the
\textit{ab}  plane as a function of pressure an apparent decrease
of \textit{H$_{int}$} would be also observed \cite{MitoT:Pre-in}.
In such a scenario, additional magnetic diffraction peaks
corresponding to a propagation vector different from
(0.5,0.5,$\delta$) would appear and a subsequent reduction of the
in-plane component would be observed. We did not observe a large
reduction of the in-plane component nor any evidence of magnetic
diffraction at (0.5,0.5,0.5) due to an AFM out of the \textit{a-b}
plane component but we cannot discardrule out magnetic intensity
appearing at (0.5,0.5,0). Taken together, our results would seem
to rule out the canting scenario. An alternative, and more
plausible, interpretation of the reduction of \textit{H$_{int}$}
is that hyperfine coupling decreases with pressure
\cite{WeiBao:pressure}. Irrespective of the magnitude of
\textit{H$_{int}$}, NQR measurements \cite{MitoT:cond} establish
beyond reasonable doubt the coexistence of AFM and bulk SC in
CeRhIn$_{5}$ at 1.75~GPa. Our diffraction results indicate that
\textit{M$_{Q}$}$\simeq$ (0.67~$\pm$~0.04)~$\mu_{B}$/Ce at 1.6~GPa
and 1.85~K. These results indicate that bulk SC coexists with
relatively large-moment AFM order in
CeRhIn$_{5}$ under pressure.\\
Unlike UPd$_{2}$Al$_{3}$ \cite{SatoNK:Strcbl} where the
coexistence of AFM and unconventional SC has been ascribed to the
partition of the three U 5\textit{f} electrons into dual roles,
magnetic and SC, CeRhIn$_{5}$ has only a single 4\textit{f}
electron that participates in creating both states. This situation
in CeRhIn$_{5}$ is also distinctly different from other
pressure-induced HFS based on Ce. CeIn$_{3}$, on which
CeRhIn$_{5}$ is based is an example. In CeIn$_{3}$ the ordered
moment \cite{MorinP:Magsup} and specific heat anomaly at
\textrm{T}$_{\textrm{N}}~$ decrease monotonically towards zero as
the critical pressure is approached where SC
appears\cite{KnebelG:press}. We do not understand presently how
such a large-moment AFM can coexist with unconventional SC in
CeRhIn$_{5}$. It is as if the 4\textit{f} moments, in some way,
also assumed dual character, perhaps purely in a dynamically as
suggested by recent NMR studies \cite{CurroNJ:cond} or spatially
segregating into AFM and SC domains. Such segregation, however,
also could be dynamic since there is no evidence for additional
NQR frequencies\cite{MitoT:cond, MitoT:SysNst, MitoT:Pre-in}.\par
In summary, we have determined a P-T phase diagram from high
quality CeRhIn$_{5}$ single crystals which shows a broad range of
pressures where AFM and SC coexist. In addition, our single
crystal magnetic neutron diffraction studies on CeRhIn$_{5}$ find
only small changes in the incommensurate magnetic structure and
ordered moment as pressure is increased up to 1.63~GPa. These
results are consistent with specific heat measurements but
inconsistent with estimates of \textit{H$_{int}$} determined by
NQR, which we attribute tentatively to a pressure-induced change
in the hyperfine coupling. We have not reproduced the observation
of a significant change in $\delta$ and the absence of AFM at
1.3~GPa reported earlier \cite{MajumdarS:pre}. Most importantly,
we have found that
 compared to other heavy fermions, the relationship between AFM and
 unconventional SC is qualitatively different in
 CeRhIn$_{5}$ and will require  the development of a new
 interpretative framework in which the 4\textit{f} electron
 produces both long-range AFM order and heavy-quasiparticles that
 pair to form an unconventional SC state.\par

Work at Los Alamos was performed under the auspices of the U.S.
Department of Energy. We would like to thank D. Pines, Ar. Abanov,
Z. Fisk, N. J. Curro and S. Nakatsuji for fruitful discussions. A.
Cull, I. Swainson, Th. Beaufils and J.-L. Meuriot are also
acknowledged for their assistance with the experiments at CRL and
LLB.  The experiments of E.G.M. at LLB were supported by the
European Commission under the Access to Research Infrastructure
Action of the Improving Human potential Programme (contract
HRPI-CT-1999-0032). \par

\noindent $*$ Electronic address: allobet$@$lanl.gov\\
 \noindent
 $\dag$ Present address:Physics Department, Brookhaven National
Laboratory, Upton, New York 11973 \& NIST Center for Neutron
Research, National Institute of Standards and Technology,
Gaithersburg, Maryland 20899-8562\\
 \noindent
 $\ddag$ Present address:Max Planck Institute for
Chemical Physics of Solids, N\"{o}thnitzer Str. 40, 01187 Dresden,
Germany \\
 \noindent $\S$ Present address: Instituto de
F\'{\i}sica ``Gleb Wataghin'', UNICAMP, 13083-970, Campinas-SP,
Brazil
 

\end{document}